\documentclass[showpacs,amsmath,amssymb,10pt]{article}
\usepackage{graphicx,color}
\usepackage{indentfirst}
\usepackage{multicol}
\usepackage{mathrsfs}

\begin{document}

\begin{flushleft}
\hspace{0.5cm}{\Large{\textbf{The Feynman rules for neutrinos and  new neutralinos in BLMSSM\footnote{supported by the Major Project of
NNSFC (No. 11535002) and NNSFC (No. 11275036),
the Natural Science Foundation of Hebei province with Grant
No. A2016201010, and the Foundation of Hebei province
with Grant No. BR2-201, and the Natural Science Fund of
Hebei University with Grants No. 2011JQ05 and No. 2012-
242, Hebei Key Lab of Optic-Electronic Information and
Materials, the midwest universities comprehensive strength
promotion project}}}}\\
\hspace{3cm}Xing-Xing Dong $^1$\footnote{dxx-0304@163.com}, Shu-Min Zhao $^{1,2}$\footnote{zhaosm@hbu.edu.cn}, Hai-Bin Zhang $^{1}$,\\
\hspace{5cm} Fang Wang $^{3}$, Tai-Fu Feng $^{1,2}$ \\
\hspace{2cm}$^1$Department of Physics and Technology, Hebei University,
Baoding 071002, China.\\
$^2$State Key Laboratory of Theoretical Physics (KLTP),
Institute of theoretical Physics, Chinese Academy of Sciences.\\
\hspace{1cm}$^3$Department of Electronic and Information Engineering, Hebei University, Baoding 071002, China.

\begin{flushleft}
\hspace{0.7cm}\textbf{Abstract} In a supersymmetric extension of the standard model where baryon and lepton numbers are local \\\hspace{0.7cm}gauge symmetries(BLMSSM),
 we deduce the Feynman rules for neutrinos and new neutralinos. We briefly \\\hspace{0.7cm}introduce the mass matrices for the particles and the related couplings in this work, which are very useful to \\\hspace{0.7cm}research the neutrinos and new neutralinos.\\
\end{flushleft}
\hspace{0.7cm}\textbf{PACS number:} {13.15.+g, 12.60.-y} \\
\hspace{0.7cm}\textbf{Key words:} {Supersymmetry, Feynman rules, mass matrices}
\end{flushleft}

\begin{multicols}{2}

\section{Introduction}
In the quantum field theory, the Standard Model (SM) is a theory concerning the electromagnetic weak and strong interactions.
Though the lightest CP-even Higgs($m_{h^0}\simeq126$ GeV) was detected by LHC,
SM makes some phenomena unexplained such as falls short of being a complete theory of fundamental interactions.
In neutrino sector, the observations of solar and atmospheric neutrino oscillations
  \cite{neutrino1,neutrino2,neutrino3,neutrino4} are not incorporated in SM, which
provides clear evidence for physics beyond SM. Furthermore, the authors think that the well-motivated dark matter candidate emerges
 from neutrino sector\cite{dark matter1,dark matter2,dark matter3}.

The physics beyond the SM has drawn the physicists' attentions for a long time. One of the appealing theories to describe physics at the TeV scale is the minimal supersymmetric extension of the standard model (MSSM)\cite{MSSM1,MSSM2,MSSM3,MSSM4}. MSSM includes the necessary additional new particles that are able to be superparters of those in SM. The right-handed neutrino superfields can extend the next to minimal supersymmetric standard model (NMSSM), and these superfields only couple with the singlet Higgs\cite{NMSSM1,NMSSM2,NMSSM3,NMSSM4}. In R-parity\cite{R-parity} conserved MSSM, the left-handed light neutrinos are still massless leading to the failure to explain
 the discovery from neutrino oscillations. Therefore, theoretical physicists extend MSSM to account for the light neutrino masses and mixings.

As the extension of the MSSM considered the local gauged baryon (B) and lepton (L) symmetries, BLMSSM is spontaneously broken at the TeV scale\cite{BLMSSM1,BLMSSM2,SU1,SU2}. In BLMSSM, the lepton number is broken in an even number while baryon number can be changed by baryon number violating operators through one unit. BLMSSM can not only account for the asymmetry of matter-antimatter in the universe but also explain the data from neutrino oscillation experiments\cite{experiments1,experiments2,experiments3}. Compared with MSSM, BLMSSM includes many new fields such as the new quarks, new leptons, new Higgs, the superfields $\hat{X}$ and $\hat{X'}$\cite{SMBL1,SMBL2,SMBL3}. In this work, we mainly study the Feynman rules for the neutrino and new neutralinos in BLMSSM.

In BLMSSM, the light neutrinos get masses from the seesaw mechanism, and proton decay is forbidden\cite{BLMSSM1,BLMSSM2,SU1,SU2}. Therefore, it is not necessary to build a large desert between the electroweak scale and grand unified scale. This is the main motivation for the BLMSSM. Many possible signals of the MSSM at the LHC have been studied by the experiments. However, with the broken B and L symmetries, the predictions and bounds for the collider experiments should be changed. From the decays of right handed neutrinos\cite{SU1,SU2,LHC}, we can look for lepton number violation at the LHC.
 Similarly from the decays of squarks and gauginos, we can also detect the baryon number violation at the LHC.
 For example, the channels with multi-tops and multi-bottoms may be caused by the baryon number violating decays of gluinos\cite{SU1,SU2}.

After this introduction, we briefly summarize the dominate contents of BLMSSM in section 2. The mass matrices for the particles are collected is section 3. Sections 4 and 5 are respectively devoted to the related couplings of neutralinos and neutrinos beyond MSSM. We show our discussion and conclusion in section 6.
\section{Some content of BLMSSM}
  Extending the MSSM with local gauged baryon $(B)$ and lepton $(L)$ numbers, one obtains
  BLMSSM, and at the TeV scale the local gauge symmetries are spontaneously broken.
  In this section, we briefly review some features of the BLMSSM.  $SU(3)_C\otimes{SU(2)_L}\otimes{U(1)_Y}\otimes{U(1)_B}\otimes{U(1)_L}$\cite{SU1,SU2} is the basic gauge symmetry of BLMSSM. The exotic leptons $\widehat{L}_4\sim(1,2,-1/2,0,L_4)$, $\widehat{E}_4^c\sim(1,1,1,0,-L_4)$, $\widehat{N}_4^c\sim(1,1,0,0,-L_4)$, $\widehat{L}_5^c\sim(1,2,1/2,0,-(3+L_4))$, $\widehat{E}_5\sim(1,1-1,0,3+L_4)$, $\widehat{N}_5\sim(1,1,0,0,3+L_4)$ and the exotic quarks $\widehat{Q}_4\sim(3,2,1/6,B_4,0)$, $\widehat{U}_4^c\sim (\overline{3},1,-2/3,-B_4,0)$, $\widehat{D}_4^c\sim(\overline{3},1,1/3,-B_4,0)$, $\widehat{Q}_5^c\sim(\overline{3},2,-1/6,-(1+B_4),0)$, $\widehat{U}_5\sim (3,1,2/3,1+B_4,0)$, $\widehat{D}_5\sim(3,1-1/3,1+B_4,0)$ are introduced to cancle L and B anomalies respectively.
  The exotic Higgs superfields $\widehat{\Phi}_L\sim(1,1,0,0,-2)$, $\widehat{\varphi}_L\sim(1,1,0,0,2)$ and $\widehat{\Phi}_B\sim(1,1,0,1,0)$, $\widehat{\varphi}_B\sim(1,1,0,-1,0)$ are introduced respectively to break lepton number and baryon number spontaneously. The exotic Higgs superfields $\widehat{\Phi}_L$, $\widehat{\varphi}_L$ and $\widehat{\Phi}_B$, $\widehat{\varphi}_B$ acquire nonzero vacuum expectation values (VEVs), then the exotic leptons and exotic quarks obtain masses. The model also includes the superfields
  $\widehat{X}\sim(1,1,0,2/3+B_4,0)$ and $\widehat{X}^{'}\sim(1,1,0,-(2/3+B_4),0)$ to make heavy exotic quarks unstable. Furthermore, the lightest mass eigenstate can be a dark matter candidate, while $\widehat{X}$ and $\widehat{X}^{'}$ mix together. Anomaly cancellation
   requires the emergence of new families. However there is no flavour violation at
tree level since they do not mix with the SM fermions and there are no Landau poles at the low scale due to the new families.

The superpotential of BLMSSM is given by \cite{superpotential}
\begin{eqnarray}
&&{\cal W}_{BLMSSM}
={\cal W}_{MSSM}+{\cal W}_{B}+{\cal W}_{L}+{\cal W}_{X},
\end{eqnarray}
where ${\cal W}_{MSSM}$ represents the superpotential of the MSSM. The concrete forms of ${\cal W}_{B}$, ${\cal W}_{L}$ and ${\cal W}_{X}$ read as follows
\begin{eqnarray}
&&{\cal W}_{B}=\lambda_{Q}\hat{Q}_{4}\hat{Q}_{5}^c\hat{\Phi}_{B}+\lambda_{U}\hat{U}_{4}^c\hat{U}_{5}
\hat{\varphi}_{B}+\lambda_{D}\hat{D}_{4}^c\hat{D}_{5}\hat{\varphi}_{B}\nonumber\\&&\hspace{1cm}+\mu_{B}\hat{\Phi}_{B}\hat{\varphi}_{B}+Y_{{u_4}}\hat{Q}_{4}\hat{H}_{u}\hat{U}_{4}^c+Y_{{d_4}}\hat{Q}_{4}\hat{H}_{d}\hat{D}_{4}^c
\nonumber\\&&\hspace{1cm}+Y_{{u_5}}\hat{Q}_{5}^c\hat{H}_{d}\hat{U}_{5}+Y_{{d_5}}\hat{Q}_{5}^c\hat{H}_{u}\hat{D}_{5},
\nonumber\\&&{\cal W}_{L}=Y_{{e_4}}\hat{L}_{4}\hat{H}_{d}\hat{E}_{4}^c+Y_{{\nu_4}}\hat{L}_{4}\hat{H}_{u}\hat{N}_{4}^c
+Y_{{e_5}}\hat{L}_{5}^c\hat{H}_{u}\hat{E}_{5}\nonumber\\&&\hspace{1cm}+Y_{{\nu_5}}\hat{L}_{5}^c\hat{H}_{d}\hat{N}_{5}
+Y_{\nu}\hat{L}\hat{H}_{u}\hat{N}^c+\lambda_{{N^c}}\hat{N}^c\hat{N}^c\hat{\varphi}_{L}
\nonumber\\&&\hspace{1cm}+\mu_{L}\hat{\Phi}_{L}\hat{\varphi}_{L},
\nonumber\\&&{\cal W}_{X}=\lambda_1\hat{Q}\hat{Q}_{5}^c\hat{X}+\lambda_2\hat{U}^c\hat{U}_{5}\hat{X}^\prime
\nonumber\\&&\hspace{1cm}+\lambda_3\hat{D}^c\hat{D}_{5}\hat{X}^\prime+\mu_{X}\hat{X}\hat{X}^\prime. \label{SWBLX}
\end{eqnarray}

The soft breaking terms $\mathcal{L}_{soft}$ of the BLMSSM are generally shown as\cite{BLMSSM1,BLMSSM2,superpotential}
\begin{eqnarray}
&&{\cal L}_{{soft}}={\cal L}_{{soft}}^{MSSM}-(m_{{\tilde{\nu}^c}}^2)_{{IJ}}\tilde{N}_I^{c*}\tilde{N}_J^c
-m_{{\tilde{Q}_4}}^2\tilde{Q}_{4}^\dagger\tilde{Q}_{4}\nonumber\\&&\hspace{0.4cm}-m_{{\tilde{U}_4}}^2\tilde{U}_{4}^{c*}\tilde{U}_{4}^c-m_{{\tilde{D}_4}}^2\tilde{D}_{4}^{c*}\tilde{D}_{4}^c
-m_{{\tilde{Q}_5}}^2\tilde{Q}_{5}^{c\dagger}\tilde{Q}_{5}^c\nonumber\\&&\hspace{0.4cm}-m_{{\tilde{U}_5}}^2\tilde{U}_{5}^*\tilde{U}_{5}
-m_{{\tilde{D}_5}}^2\tilde{D}_{5}^*\tilde{D}_{5}
-m_{{\tilde{L}_4}}^2\tilde{L}_{4}^\dagger\tilde{L}_{4}\nonumber\\&&\hspace{0.4cm}-m_{{\tilde{\nu}_4}}^2\tilde{N}_{4}^{c*}\tilde{N}_{4}^c
-m_{{\tilde{e}_4}}^2\tilde{E}_{_4}^{c*}\tilde{E}_{4}^c
-m_{{\tilde{L}_5}}^2\tilde{L}_{5}^{c\dagger}\tilde{L}_{5}^c
\nonumber\\&&\hspace{0.4cm}-m_{{\tilde{\nu}_5}}^2\tilde{N}_{5}^*\tilde{N}_{5}-m_{{\tilde{e}_5}}^2\tilde{E}_{5}^*\tilde{E}_{5}-m_{{\Phi_{B}}}^2\Phi_{B}^*\Phi_{B}
\nonumber\\&&\hspace{0.4cm}-m_{{\varphi_{B}}}^2\varphi_{B}^*\varphi_{B}-m_{{\Phi_{L}}}^2\Phi_{L}^*\Phi_{L}
-m_{{\varphi_{L}}}^2\varphi_{L}^*\varphi_{L}\nonumber\\&&\hspace{0.4cm}-\Big(m_{B}\lambda_{B}\lambda_{B}+m_{L}\lambda_{L}\lambda_{L}+h.c.\Big)
\nonumber\\&&\hspace{0.4cm}+\Big\{A_{{u_4}}Y_{{u_4}}\tilde{Q}_{4}H_{u}\tilde{U}_{4}^c+A_{{d_4}}Y_{{d_4}}\tilde{Q}_{4}H_{d}\tilde{D}_{4}^c
\nonumber\\&&\hspace{0.4cm}+A_{{u_5}}Y_{{u_5}}\tilde{Q}_{5}^cH_{d}\tilde{U}_{5}+A_{{d_5}}Y_{{d_5}}\tilde{Q}_{5}^cH_{u}\tilde{D}_{5}
\nonumber\\&&\hspace{0.4cm}+A_{{BQ}}\lambda_{Q}\tilde{Q}_{4}\tilde{Q}_{5}^c\Phi_{B}+A_{{BU}}\lambda_{U}\tilde{U}_{4}^c\tilde{U}_{5}\varphi_{B}
\nonumber\\&&\hspace{0.4cm}+A_{{BD}}\lambda_{D}\tilde{D}_{4}^c\tilde{D}_{5}\varphi_{B}+B_{B}\mu_{B}\Phi_{B}\varphi_{B}+h.c.\Big\}
\nonumber\\&&\hspace{0.4cm}+\Big\{A_{{e_4}}Y_{{e_4}}\tilde{L}_{4}H_{d}\tilde{E}_{4}^c+A_{{\nu_4}}Y_{{\nu_4}}\tilde{L}_{4}H_{u}\tilde{N}_{4}^c
\nonumber\\&&\hspace{0.4cm}+A_{{e_5}}Y_{{e_5}}\tilde{L}_{5}^cH_{u}\tilde{E}_{5}+A_{{\nu_5}}Y_{{\nu_5}}\tilde{L}_{5}^cH_{d}\tilde{N}_{5}
\nonumber\\&&\hspace{0.4cm}+A_{\nu}Y_{\nu}\tilde{L}H_{u}\tilde{N}^c+A_{{\nu^c}}\lambda_{{\nu^c}}\tilde{N}^c\tilde{N}^c\varphi_{L}
\nonumber\\&&\hspace{0.4cm}+B_{L}\mu_{L}\Phi_{L}\varphi_{L}+h.c.\Big\}
\nonumber\\&&\hspace{0.4cm}+\Big\{A_1\lambda_1\tilde{Q}\tilde{Q}_{5}^cX+A_2\lambda_2\tilde{U}^c\tilde{U}_{5}X^\prime
\nonumber\\&&\hspace{0.4cm}+A_3\lambda_3\tilde{D}^c\tilde{D}_{5}X^\prime+B_{X}\mu_{X}XX^\prime+h.c.\Big\},\label{soft}
\end{eqnarray}
where ${\cal L}_{{soft}}^{MSSM}$ represent the soft breaking terms of MSSM, $\lambda_{B}$ and $\lambda_{L}$ are the gauginos of $U(1)_B$ and $U(1)_L$, respectively.  The $SU(2)_L$ doublets $H_{u},\;H_{d}$ and $SU(2)_L$ singlets $\Phi_{B},\;\varphi_{B},\;\Phi_{L},\;
\varphi_{L}$ acquire the nonzero VEVs $\upsilon_{u},\;\upsilon_{d}$ and $\upsilon_{{B}},\;\overline{\upsilon}_{{B}},\;\upsilon_{L},\;\overline{\upsilon}_{L}$ respectively,
\begin{eqnarray}
&&H_u=\left({\begin{array}{*{20}{c}}
H_u^+  \\
\frac{1}{\sqrt 2}(\upsilon_u+H_u^0+iP_u^0)  \\
\end{array}}
\right), \nonumber\\
&&H_d=\left({\begin{array}{*{20}{c}}
\frac{1}{\sqrt 2}(\upsilon_d+H_d^0+iP_d^0)  \\
H_d^-  \\
\end{array}}
\right),\nonumber\\
&&\Phi_B=\frac{1}{\sqrt{2}} (\upsilon_B+\Phi_B^0+iP_B^0), \nonumber\\
&&\varphi_B=\frac{1}{\sqrt{2}} (\overline{\upsilon}_B+\varphi_B^0+i\overline{P}_B^0),\nonumber\\
&&\Phi_L=\frac{1}{\sqrt{2}}(\upsilon_L+\Phi_L^0+iP_L^0),\nonumber\\
&&\varphi_L=\frac{1}{\sqrt{2}}
(\overline{\upsilon}_L+\varphi_L^0+i\overline{P}_L^0).
\end{eqnarray}
Therefore, the local gauge symmetry
$SU(2)_L\otimes{U(1)_Y}\otimes{U(1)_B}\otimes{U(1)_L}$ is broken down to the electromagnetic symmetry $U(1)_e$.
In Ref.\cite{superpotential}, the mass matrices of exotic Higgs, exotic quarks and exotic scalar quarks are obtained.
In BLMSSM, because of the introduced superfields $\hat{N}^C$, the tiny masses of the light neutrinos are produced.
Another result is six scalar neutrinos in BLMSSM.
\section{the mass matrices for the particles}
Lepton neutralinos are made up of $\lambda_L$ (the superpartner of the new lepton boson), $\psi_{\Phi_L}$ and $\psi_{\varphi_L}$
 (the superpartners of the $SU(2)_L$ singlets $\Phi_L$ and $\varphi_L$). The mixing mass matrix of lepton neutralinos is shown in the
 basis $(i\lambda_L,\psi_{\Phi_L},\psi_{\varphi_L})$ \cite{lepton neutralinos1,lepton neutralinos2}. Then 3 lepton
  neutralino masses are obtained from diagonalizing the mass mixing matrix $M_{LN}$ by $Z_{N_L}$,
 \begin{eqnarray}
&& M_{LN}=\left( \begin{array}{ccc}
  2M_L &2v_Lg_L &-2\bar{v}_Lg_L\\
   2v_Lg_L & 0 &-\mu_L\\-2\bar{v}_Lg_L&-\mu_L &0
    \end{array}\right),
    \nonumber\\&& i\lambda_L=Z_{N_L}^{1i}k_{L_i}^0,~~
   \psi_{\Phi_L}=Z_{N_L}^{2i}k_{L_i}^0,~~\nonumber\\&&
   \psi_{\varphi_L}=Z_{N_L}^{3i}k_{L_i}^0,~~ \chi^0_{L_i}= \left(\begin{array}{c}
 k_{L_i}^0\\ \bar{k}_{L_i}^0
    \end{array}\right).
  \end{eqnarray}
$\chi^0_{L_i} (i=1,2,3)$ are the mass eigenstates of lepton neutralinos.

 $\lambda_B$ (the superpartner of the new baryon boson), $\psi_{\Phi_B}$ and $\psi_{\varphi_B}$
 (the superpartners of the $SU(2)_L$ singlets $\Phi_B$ and $\varphi_B$) mix together producing 3 baryon neutralinos.
 Using $Z_{N_B}$ one can diagonalize the mass mixing matrix $M_{BN}$, and obtain 3 baryon neutralino masses,
 \begin{eqnarray}
&&M_{BN}=\left(  \begin{array}{ccc}
  2M_B &-v_Bg_B & \bar{v}_Bg_B\\-
   v_Bg_B & 0 &-\mu_B\\ \bar{v}_Bg_B&-\mu_B &0
    \end{array}\right).
     \nonumber\\&& i\lambda_B=Z_{N_B}^{1i}k_{B_i}^0,~~
   \psi_{\Phi_B}=Z_{N_B}^{2i}k_{B_i}^0,~~\nonumber\\
   &&\psi_{\varphi_B}=Z_{N_B}^{3i}k_{B_i}^0,~~ \chi^0_{B_i}= \left(\begin{array}{c}
 k_{B_i}^0\\ \bar{k}_{B_i}^0
    \end{array}\right).
   \end{eqnarray}
  The mass eigenstates of baryon neutralinos are represented by $\chi^0_{B_i} (i=1,2,3)$.

  In the work, because neutrinos are majorana particles, we can use the following expression.
In the base $( \psi_ {\nu^I_L},\psi_{N^{cI}_R} )$, the formulas for neutrino mixing and mass matrix are
shown as

\begin{eqnarray}
&&Z_{N_\nu}^{T}\left(\begin{array}{cc}
  0&\frac{v_u}{\sqrt{2}}(Y_{\nu})^{IJ} \\
   \frac{v_u}{\sqrt{2}}(Y^{T}_{\nu})^{IJ}  & \frac{\bar{v}_L}{\sqrt{2}}(\lambda_{N^c})^{IJ}
    \end{array}\right) Z_{N_\nu}
  \nonumber\\&&=diag(m_{\nu^\alpha}), ~~~~ \alpha=1\dots 6.
    \nonumber\\&& \psi_{\nu^I_L}=Z_{N_\nu}^{I\alpha}k_{N_\alpha}^0,~~~~
   \psi_{N^{cI}_R}=Z_{N_\nu}^{(I+3)\alpha}k_{N_\alpha}^0,~~~~\nonumber\\
   &&\chi_{N_\alpha}^0= \left(\begin{array}{c}
 k_{N_\alpha}^0\\ \bar{k}_{N_\alpha}^0
    \end{array}\right).\label{neutrinoD}
      \end{eqnarray}

 $\chi_{N_\alpha}^0$ denote the mass eigenstates of the neutrino fields mixed by the left-handed and right-handed neutrinos.

The introduced super-fields $\hat{N}^c$ lead to six sneutrinos. From the superpotential and the soft breaking terms in Eqs.(\ref{SWBLX},\ref{soft}), we deduce the mass squared matrix of sneutrino (${\cal M}_{\tilde{n}}$) in the base $\tilde{n}^{T}=(\tilde{\nu},\tilde{N}^{c})$. To obtain mass eigenstates of sneutrinos, $Z_{\tilde{\nu}}$ is used for the rotation.
\begin{eqnarray}
  && {\cal M}^2_{\tilde{n}}(\tilde{\nu}_{I}^*\tilde{\nu}_{J})=\frac{g_1^2+g_2^2}{8}(v_d^2-v_u^2)\delta_{IJ}+g_L^2(\overline{v}^2_L\nonumber\\&&\hspace{2cm}-v^2_L)\delta_{IJ}
   +\frac{v_u^2}{2}(Y^\dag_{\nu}Y_\nu)_{IJ}+(m^2_{\tilde{L}})_{IJ},\nonumber\\&&
   {\cal M}^2_{\tilde{n}}(\tilde{N}_I^{c*}\tilde{N}_J^c)=-g_L^2(\overline{v}^2_L-v^2_L)\delta_{IJ}
   +\frac{v_u^2}{2}(Y^\dag_{\nu}Y_\nu)_{IJ}\nonumber\\&&\hspace{1.5cm}+2\overline{v}^2_L(\lambda_{N^c}^\dag\lambda_{N^c})_{IJ}
   +(m^2_{\tilde{N}^c})_{IJ}\nonumber\\&&\hspace{1.5cm}+\mu_L\frac{v_L}{\sqrt{2}}(\lambda_{N^c})_{IJ}
   -\frac{\overline{v}_L}{\sqrt{2}}(A_{N^c})_{IJ}(\lambda_{N^c})_{IJ},\nonumber\\&&
   {\cal M}^2_{\tilde{n}}(\tilde{\nu}_I\tilde{N}_J^c)=\mu^*\frac{v_d}{\sqrt{2}}(Y_{\nu})_{IJ}-v_u\overline{v}_L(Y_{\nu}^\dag\lambda_{N^c})_{IJ}
   \nonumber\\&&\hspace{2.2cm}+\frac{v_u}{\sqrt{2}}(A_{N})_{IJ}(Y_\nu)_{IJ}.\nonumber\\&&
  Z_{\tilde{\nu}}^\dag{\cal M}_{\tilde{n}}Z_{\tilde{\nu}}=(m_{\tilde{N}^{1}}^2,m_{\tilde{N}^{2}}^2,m_{\tilde{N}^{3}}^2
,m_{\tilde{N}^{4}}^2,m_{\tilde{N}^{5}}^2,m_{\tilde{N}^{6}}^2). \label{SneutrinoD}
   \end{eqnarray}

The superfields $\Phi_{B}^0$ and $\varphi_{B}^0$ mix together, whose mass squared matrix is
\end{multicols}
\begin{eqnarray}
&&{\cal M}_{{EB}}^2=\left(\begin{array}{ll}m_{{Z_B}}^2\cos^2\beta_{B}+m_{{A_{B}^0}}^2\sin^2\beta_{B},\;&
(m_{{Z_B}}^2+m_{{A_{B}^0}}^2)\cos\beta_{B}\sin\beta_{B}\\
(m_{{Z_B}}^2+m_{{A_{B}^0}}^2)\cos\beta_{B}\sin\beta_{B},\;&
m_{{Z_B}}^2\sin^2\beta_{B}+m_{{A_{B}^0}}^2\cos^2\beta_{B}
\end{array}\right),
\nonumber\\&&v_{B_t}=\sqrt{\upsilon_{B}^2+\overline{\upsilon}_{B}^2},~~~m_{{Z_B}}=g_{B}v_{B_t},
~~~\Phi_B^0=Z^{1i}_{\phi_B}H^0_{B_i},~~~\varphi_B^0=Z^{2i}_{\phi_B}H^0_{B_i},
\label{CPevenB-mass}
\end{eqnarray}
with $m_{{Z_B}}$ representing the mass of
neutral $U(1)_{B}$ gauge boson $Z_{B}$.
$Z_{\phi_B}$ is the rotation matrix to diagonalize the mass squared matrix ${\cal M}_{{EB}}^2$ and
$H^0_{B_i} (i=1,2)$ denote the mass eigenstates of baryon Higgs.

In the same way, we obtain the mass squared matrix for $(\Phi_{L}^0,\;\varphi_{L}^0)$
\begin{eqnarray}
&&{\cal M}_{{EL}}^2=\left(\begin{array}{ll}m_{{Z_L}}^2\cos^2\beta_{_L}+m_{{A_{L}^0}}^2\sin^2\beta_{L},\;&
(m_{{Z_L}}^2+m_{{A_{L}^0}}^2)\cos\beta_{L}\sin\beta_{L}\\
(m_{{Z_L}}^2+m_{{A_{L}^0}}^2)\cos\beta_{L}\sin\beta_{L},\;&
m_{{Z_L}}^2\sin^2\beta_{L}+m_{{A_{L}^0}}^2\cos^2\beta_{L}
\end{array}\right),
\nonumber\\&&
v_{L_t}=\sqrt{\upsilon_{_L}^2+\overline{\upsilon}_{L}^2},~~~m_{{Z_L}}=2g_{L}v_{L_t},
~~~\Phi_L^0=Z^{1i}_{\phi_L}H^0_{L_i},~~~\varphi_L^0=Z^{2i}_{\phi_L}H^0_{L_i}.
\label{CPevenL-mass}
\end{eqnarray}
\begin{multicols}{2}
Here, $m_{{Z_L}}$ is the mass of
neutral $U(1)_{L}$ gauge boson $Z_{L}$. $Z_{\phi_L}$ is used to obtain mass eigenvalues for the matrix ${\cal M}_{_{EL}}^2$.
$H^0_{L_i} (i=1,2)$ are the lepton Higgs mass eigenstates.
\section{the couplings of neutralinos beyond MSSM}
\subsection{the new couplings of MSSM neutralinos}
From the superpotential $\mathcal{W}_L$ in Eq.(\ref{SWBLX}) and the interactions of gauge and matter multiplets
$ig\sqrt{2}T^a_{ij}(\lambda^a\psi_jA_i^*-\bar{\lambda}^a\bar{\psi}_iA_j)$, we deduce
 the couplings of MSSM neutralino-exotic lepton-exotic slepton
\begin{eqnarray}
&&\mathcal{L}(\chi^0l'\tilde{l}')=\sum_{i,k=1}^2\sum_{j=1}^4\Big\{
\bar{\chi}_j^0\Big[\Big(\frac{1}{\sqrt{2}}(\frac{e}{s}Z_N^{2j}\nonumber\\&&\hspace{0.4cm}+\frac{e}{c}Z_N^{1j})U_L^{1i}Z_{\tilde{e}_4}^{1k*}+
Y_{e_4}U_L^{1i}Z_N^{3j}Z_{\tilde{e}_4}^{2k*}\Big)P_L\nonumber\\&&\hspace{0.4cm}+\Big(Y_{e_4}^*Z_{\tilde{e}_4}^{1k*}Z_N^{3j*}W_L^{2i}
\nonumber\\&&\hspace{0.4cm}-\sqrt{2}\frac{e}{c}Z_N^{1j*}W_L^{2i}Z_{\tilde{e}_4}^{2k*}\Big)P_R\Big]L'_{i+3}\tilde{E}_4^{k*}
\nonumber\\&&\hspace{0.4cm}-\bar{\chi}_j^0\Big[\Big(Y_{\nu_4}U_N^{1i}Z_N^{4j}Z_{\tilde{\nu}_4}^{2k*}\nonumber\\&&\hspace{0.4cm}+\frac{1}{\sqrt{2}}(\frac{e}{s}Z_N^{2j}-
\frac{e}{c}Z_N^{1j})U_N^{1i}Z_{\tilde{\nu}_4}^{1k*}\Big)P_L
\nonumber\\&&\hspace{0.4cm}+Y_{\nu_4}^*Z_{\tilde{\nu}_4}^{1k*}Z_N^{4j*}W_N^{2i}P_R\Big]N'_{i+3}\tilde{N}_4^{k*}
\nonumber\\&&\hspace{0.4cm}-\bar{L'}_{i+3}\Big[\Big(Y_{e_5}Z_N^{4j}W_L^{1i*}Z_{\tilde{e}_5}^{1k}
\nonumber\\&&\hspace{0.4cm}-\frac{1}{\sqrt{2}}(\frac{e}{s}Z_N^{2j}+\frac{e}{c}Z_N^{1j})W_L^{1i*}Z_{\tilde{e}_5}^{2k}\Big)P_L
\nonumber\\&&\hspace{0.4cm}+\Big(Y_{e_5}^*Z_{\tilde{e}_5}^{1k}Z_N^{4j*}U_L^{2i*}\nonumber\\&&\hspace{0.4cm}+\sqrt{2}\frac{e}{c}Z_N^{1j*}U_L^{2i*}Z_{\tilde{e}_5}^{1k}\Big)P_R\Big]\chi_j^0\tilde{E}_5^k
\nonumber\\&&\hspace{0.4cm}+\bar{N'}_{i+3}\Big[\Big(Y_{\nu_5}W_N^{1i*}Z_N^{3j}Z_{\tilde{\nu}_5}^{1k}
\nonumber\\&&\hspace{0.4cm}+\frac{1}{\sqrt{2}}(\frac{e}{s}Z_N^{2j}-\frac{e}{c}Z_N^{1j})W_N^{1i*}Z_{\tilde{\nu}_5}^{2k}\Big)P_L
\nonumber\\&&\hspace{0.4cm}+Y_{\nu_5}^*Z_{\tilde{\nu}_5}^{2k}Z_N^{3j*}U_N^{2i*}P_R\Big]\chi_j^0\tilde{N}_5^k\Big\}+H.c.
\end{eqnarray}
The matrices $U_L$ and $W_L$ are used to diagonalize the exotic charged lepton mixing matrix\cite{SMBL1,SMBL2,SMBL3},
and $L'_{4,5}$ are the mass eigenstates of the exotic charged leptons. While the exotic slepton mass eigenstates
are denoted by $\tilde{N}'_{4,5}$ and $\tilde{E}'_{4,5}$ with the rotation matrices $Z_{\tilde{\nu}_{4,5}}$ and $Z_{\tilde{e}_{4,5}}$. In MSSM, there are couplings for MSSM neutralino-neutrino-sneutrino which should be transformed into BLMSSM with
the rotations of the neutrinos and sneutrinos in Eqs.(\ref{neutrinoD}, \ref{SneutrinoD}).

In $\mathcal{W}_L$ there is a new term $Y_{\nu}\hat{L}\hat{H}_{u}\hat{N}^c$ that can
give corrections to the couplings of MSSM neutralino-neutrino-sneutrino. These new couplings are suppressed by the
tiny neutrino Yukawa $Y_\nu$,
\begin{eqnarray}
&&\mathcal{L}^n(\chi_{N}\chi^0\tilde{N}^{*})\nonumber\\
&&\hspace{0.2cm}=-\sum_{I,J=1}^3\;\sum_{i=1}^4\;\sum_{j=1}^6\bar{\chi}_{N_\alpha}
\Big(Y^{IJ}_{\nu}Z_{N_\nu}^{I\alpha}Z_N^{4i}Z_{\tilde{\nu}}^{(J+3)j*}P_L
\nonumber\\&&\hspace{0.6cm}+Y_{\nu}^{IJ*}Z_{\tilde{\nu}}^{Ij*}Z_N^{4i*}Z_{N_\nu}^{(J+3)\alpha*}P_R\Big)\chi_i^0\tilde{N}^{j*}+H.c.
\end{eqnarray}

In the same way, the couplings of MSSM neutralino-exotic quark-exotic squark are obtained
\begin{eqnarray}
&&\mathcal{L}(\chi^0q'\tilde{q}')=\sum_{j=1}^2\;\sum_{i,k=1}^4\Big\{
\bar{\chi}_i^0\Big[\Big(Y_{U_5}U_{3k}^*Z_N^{3i}U_{2j}^{t'}\nonumber\\&&\hspace{0.4cm}-\frac{e}{c}\frac{2\sqrt{2}}{3}Z_N^{1i}U_{2j}^{t'}U_{4k}^*
-\frac{1}{\sqrt{2}}(\frac{e}{s}Z_N^{2i}\nonumber\\&&\hspace{0.4cm}+\frac{1}{3}\frac{e}{c}Z_N^{1i})U_{1j}^{t'}U_{1k}^*
-Y_{U_4}U_{1j}^{t'}Z_N^{4i}U_{2k}^*\Big)P_L
\nonumber\\&&\hspace{0.4cm}+\Big(Y_{U_5}^*W_{1j}^{t'}Z_N^{3i*}U_{4k}^*+\frac{1}{\sqrt{2}}(\frac{e}{s}Z_N^{2i*}\nonumber\\&&\hspace{0.4cm}+\frac{1}{3}\frac{e}{c}Z_N^{1i*})W_{1j}^{t'}U_{3k}^*
-Y_{U_4}^*U_{1k}^*Z_N^{4i*}W_{2j}^{t'}
\nonumber\\&&\hspace{0.4cm}+\frac{e}{c}\frac{2\sqrt{2}}{3}Z_N^{1i*}W_{2j}^{t'}U_{2k}^*\Big)P_R\Big]t'_{j+3}\tilde{\mathcal{U}}_k^*
\nonumber\\&&\hspace{0.4cm}+\bar{\chi}_i^0[\Big(Y_{d_4}U_{1j}^{b'}Z_N^{3i}D_{2k}^*
\nonumber\\&&\hspace{0.4cm}+\frac{1}{\sqrt{2}}(\frac{e}{s}Z_N^{2i}-\frac{1}{3}\frac{e}{c}Z_N^{1i})U_{1j}^{b'}D_{1k}^*
\nonumber\\&&\hspace{0.4cm}-Y_{d_5}D_{3j}^*Z_N^{4i}U_{2k}^{b'}
+\frac{e}{c}\frac{\sqrt{2}}{3}Z_N^{1i}U_{2j}^{b'}D_{4k}^*\Big)P_L
\nonumber\\&&\hspace{0.4cm}+\Big(Y_{d_4}^*D_{1k}^*Z_N^{3i*}W_{2j}^{b'}-\frac{1}{\sqrt{2}}(\frac{e}{s}Z_N^{2i*}\nonumber\\&&\hspace{0.4cm}-\frac{1}{3}\frac{e}{c}Z_N^{1i*})W_{1j}^{b'}D_{3k}^*
-Y_{d_5}^*W_{1j}^{b'}Z_N^{4i*}D_{4k}^*
\nonumber\\&&\hspace{0.4cm}-\frac{e}{c}\frac{\sqrt{2}}{3}Z_N^{1i*}W_{2j}^{b'}D_{2k}^*\Big)P_R]b'_{j+3}\tilde{\mathcal{D}}_k^*\Big\}+H.c.
\end{eqnarray}

In the mass basis the exotic quarks are $t'$ and $b'$, whose rotation matrices are $W^{t'}, U^{t'}, W^{b'}$ and $U^{b'}$.  $\tilde{\mathcal{U}}$ and $\tilde{\mathcal{D}}$ are the exotic scalar quarks with their diagonalizing matrices  $U$ and $D$.
\subsection{the couplings of lepton neutralinos}

At tree level, lepton neutralinos not only have relations with leptons and sleptons, but also act with neutrinos and sneutrinos.
\begin{eqnarray}
&&\mathcal{L}(\chi^0_Ll\tilde{l})=\sum_{\alpha,j=1}^6\sum_{I,J,j=1}^3\bar{\chi}_{N_\alpha}\Big(
[-(\lambda_{N_c}^{IJ}\nonumber\\&&\hspace{0.4cm}+\lambda_{N_c}^{JI})Z_{N_\nu}^{(I+3)\alpha}Z_{N_L}^{3i}Z_{\tilde{\nu}}^{(J+3)j*}\nonumber\\
&&\hspace{0.4cm}+\sqrt{2}g_LZ_{N_L}^{1i}Z_{N_\nu}^{I\alpha}Z_{\tilde{\nu}}^{Jj*}\delta_{IJ}
]P_L
\nonumber\\&&\hspace{0.4cm}-\sqrt{2}g_LZ_{N_L}^{1i*}Z_{N_{\nu}}^{(I+3)\alpha*}Z_{\tilde{\nu}}^{(J+3)j*}\delta_{IJ}P_R\Big)\chi_{L_i^0}\tilde{N}^{j*}
\nonumber\\&&\hspace{0.4cm}+\sum_{i,I=1}^3\sum_{j=1}^6
\sqrt{2}g_L\bar{\chi}_{L_i^0}(Z_{N_L}^{1i}Z_{\tilde{L}}^{Ij}P_L
\nonumber\\&&\hspace{0.4cm}-Z_{N_L}^{1i*}Z_{\tilde{L}}^{(I+3)j*}P_R)e^I\tilde{L}_j^{-*}+H.c.
\end{eqnarray}
The couplings for lepton neutralino-exotic lepton-exotic slepton and
lepton neutralino-exotic neutrino-exotic sneutrino read as
\begin{eqnarray}
&&\mathcal{L}(\chi^0_Ll'\tilde{l}')
=\sum_{i=1}^3\sum_{j,k=1}^2\Big\{L_4\sqrt{2}g_L\bar{\chi}_{L_i^0}(Z_{N_L}^{1i}U_{L}^{1j}Z_{\tilde{e}_4}^{1k*}P_L
\nonumber\\&&\hspace{0.4cm}-Z_{N_L}^{1i*}Z_{\tilde{e}_4}^{2k*}W_L^{2j}P_R)L'_{j+3}\tilde{E}_4^{k*}
\nonumber\\&&\hspace{0.4cm}+L_4\sqrt{2}g_L\bar{\chi}_{L_i^0}(Z_{N_L}^{1i}U_{N}^{1j}Z_{\tilde{\nu}_4}^{1k*}P_L
\nonumber\\&&\hspace{0.4cm}-Z_{N_L}^{1i*}Z_{\tilde{\nu}_4}^{2k*}W_N^{2j}P_R)N'_{j+3}\tilde{N}_4^{k*}
\nonumber\\&&\hspace{0.4cm}+\sqrt{2}(3+L_4)g_L\bar{\chi}_{L_i}^0(Z_{N_L}^{1i}Z_{\tilde{e}_5}^{1j*}U_L^{2k}P_L
\nonumber\\&&\hspace{0.4cm}-Z_{N_L}^{1i*}W_L^{1k}Z_{\tilde{e}_5}^{2j*}P_R)L'_{k+3}\tilde{E}_5^{j*}
\nonumber\\&&\hspace{0.4cm}+\sqrt{2}(3+L_4)g_L\bar{\chi}_{L_i}^0(Z_{N_L}^{1i}Z_{\tilde{\nu}_5}^{1j*}U_N^{2k}P_L
\nonumber\\&&\hspace{0.4cm}-Z_{N_L}^{1i*}W_N^{1k}Z_{\tilde{\nu}_5}^{2j*}P_R)N'_{k+3}\tilde{N}_5^{j*}\Big\}+H.c.
\end{eqnarray}
From the interactions of gauge and matter multiplets, we write down the couplings of lepton neutralino-lepton neutralino-lepton Higgs
\begin{eqnarray}
&&\mathcal{L}(\chi_{L}^0\chi_{L}^0H_{L}^{0*})
=2\sqrt{2}g_L\sum_{i,j=1}^3\sum_{k=1}^2
Z_{N_L}^{1i}\big(Z_{N_L}^{3j}Z_{\phi_L}^{2k*}\nonumber\\
&&\hspace{0.4cm}-Z_{N_L}^{2j}Z_{\phi_L}^{1k*}\big)
\bar{\chi}_{L_i}^0P_L\chi_{L_j}^0H_{L_k}^{0*}+H.c.
\end{eqnarray}

\subsection{the couplings of baryon neutralinos}
Baryon neutralinos interact with quarks and squarks, whose couplings are in the following form
\begin{eqnarray}
&&\mathcal{L}(\chi^0_Bq\tilde{q})
=\sum_{I,i=1}^3\;\sum_{j=1}^6\Big\{
\frac{\sqrt{2}}{3}g_B\bar{\chi}_{B_i^0}(Z_{N_B}^{1i}Z_{\tilde{U}}^{Ij*}P_L\nonumber\\
&&\hspace{0.4cm}-Z_{N_B}^{1i*}Z_{\tilde{U}}^{(I+3)j*}P_R)u^I\tilde{U}_j^*
+\frac{\sqrt{2}}{3}g_B\bar{\chi}_{B_i^0}(Z_{N_B}^{1i}Z_{\tilde{D}}^{Ij}P_L
\nonumber\\&&\hspace{0.4cm}-Z_{N_B}^{1i*}Z_{\tilde{D}}^{(I+3)j*}P_R)d^I\tilde{D}_j^*\Big\}+H.c.
\end{eqnarray}
Similarly the couplings of baryon neutralino-exotic quark-exotic squark are deduced here
\begin{eqnarray}
&&\mathcal{L}(\chi^0_Bq'\tilde{q}')=\sum_{i=1}^3\sum_{j=1}^6\sum_{k=1}^2\Big\{\sqrt{2}g_B\bar{t'}_{k+3}\Big[
-\Big((1\nonumber\\&&\hspace{0.4cm}+B_4)W_{1k}^{t'*}U_{3j}Z_{N_B}^{1i}
+\lambda_QU_{1j}W_{1k}^{t'*}Z_{N_B}^{2i}
\nonumber\\&&\hspace{0.4cm}+\lambda_UW_{2k}^{t'*}Z_{N_B}^{3i}U_{4j}
+B_4U_{2j}W_{2k}^{t'*}Z_{N_B}^{1i}\Big)P_L\nonumber\\&&\hspace{0.4cm}+\Big(B_4U_{1k}^{t'*}U_{1j}Z_{N_B}^{1i*}
+(1+B_4)U_{4j}U_{2k}^{t'*}Z_{N_B}^{1i*}\nonumber\\&&\hspace{0.4cm}-\lambda_Q^*U_{1k}^{t'*}U_{3j}Z_{N_B}^{2i*}
-\lambda_U^*U_{2j}U_{2k}^{t'*}Z_{N_B}^{3i*}\Big)P_R\Big]\chi_{B_i}^0\tilde{\mathcal{U}}_j
\nonumber\\&&\hspace{0.4cm}+\sqrt{2}g_B\bar{b'}_{k+3}\Big[\Big(\lambda_QD_{1j}W_{1k}^{b'*}Z_{N_B}^{2i}\nonumber\\&&\hspace{0.4cm}-B_4D_{2j}W^{b'*}_{2k}Z_{N_B}^{1i}
-(1+B_4)W_{1k}^{b'*}D_{3j}Z_{N_B}^{1i}\nonumber\\&&\hspace{0.4cm}-\lambda_DW_{2k}^{b'*}Z_{N_B}^{3i}D_{4j}\Big)P_L
+\Big(\lambda^*_QU_{1k}^{b'*}D_{3j}Z_{N_B}^{2i*}\nonumber\\&&\hspace{0.4cm}-\lambda^*_DU_{2k}^{b'*}D_{2j}Z_{N_B}^{3i*}
+(1+B_4)D_{4j}U_{2k}^{b'*}Z_{N_B}^{1i*}\nonumber\\&&\hspace{0.4cm}+U_{1k}^{b'*}
D_{1j}Z_{N_B}^{1i*}\Big)P_R\Big]\chi_{B_i}^0\tilde{\mathcal{D}}_j\Big\}+H.c.
\end{eqnarray}
Besides the baryon neutralino-baryon neutralino-baryon Higgs couplings, there are also
interactions among baryon neutralinos and $X$ fields
\begin{eqnarray}
&&\mathcal{L}(\chi^0_B\chi^0_BH_B)=\sqrt{2}g_B\sum_{i,j=1}^3\;\sum_{k=1}^2\big(Z_{N_B}^{2j}Z_{\phi_B}^{1k*}
\nonumber\\&&\hspace{2.5cm}-Z_{N_B}^{3j}Z_{\phi_B}^{2k*}\big)Z_{N_B}^{1i}\bar{\chi}_{B_j}^0P_L\chi_{B_i}^0H_{B_k}^{0*},
\nonumber\\&&\mathcal{L}(\chi^0_B\psi_X\tilde{X})
=(\frac{2}{3}+B_4)
\sqrt{2}g_B\sum_{i=1}^3\;\sum_{j=1}^2Z_{N_B}^{1i}\bar{\chi}_{B_i}^0[Z_X^{1j*}P_L
\nonumber\\&&\hspace{2.3cm}-Z_X^{2j*}P_R]\psi_X\tilde{X}_j^*+H.c.
\end{eqnarray}

\section{the couplings of neutrino beyond MSSM}
Because of the non-zero masses and mixing angles of light neutrinos, physicists are interested in neutrino physics which implies
the lepton number violation in the Universe. In MSSM, the neutrino couplings are obtained,
so we deduce the neutrino couplings beyond MSSM in this work.
From the supperpotential $\mathcal{W }_L$ and the interactions of gauge and matter multiplets, we obtain $\mathcal{L}^1(\nu)$ and $\mathcal{L}^2(\nu)$.
$\mathcal{L}^1(\nu)$ include the neutrino couplings with Higgs: 1. neutrino-neutrino-neutral \emph{CP}-odd Higgs; 2. neutrino-neutrino-neutral \emph{CP}-even Higgs;
3. neutrino-lepton-charged Higgs; 4. neutrino-neutrino-lepton Higgs
\begin{eqnarray}
&&\mathcal{L}^1(\nu)=\sum_{I,J=1}^3\Big\{\sum_{\alpha=1}^6\sum_{j=1}^2
Y_\nu^{IJ}Z_{N_\nu}^{(J+3)\alpha}Z_H^{2j}\bar{\chi}^0_{N_\alpha}P_Le^IH_j^+
\nonumber\\&&\hspace{0.4cm}-\sum_{\alpha,\beta=1}^6\sum_{j=1}^2\frac{i}{2}Y_\nu^{IJ}Z_{N_\nu}^{I\alpha}
Z_{N_\nu}^{(3+J)\beta}Z_H^{2j}\bar{\chi}^0_{N_\beta}P_L\chi^0_{N_\alpha}A_j^0
\nonumber\\&&\hspace{0.2cm}-\sum_{\alpha,\beta=1}^6\sum_{j=1}^2\frac{1}{2}
Y_\nu^{IJ}Z_{N_\nu}^{I\alpha}Z_{N_\nu}^{(3+J)\beta}Z_R^{2j}\bar{\chi}^0_{N_\beta}P_L\chi^0_{N_\alpha}
H_j^0
\nonumber\\&&\hspace{0.2cm}-\sum_{\alpha,\beta=1}^6\sum_{i=1}^3\lambda_{N^c}^{IJ}Z_{N_\nu}^{(I+3)\alpha}
Z_{N_\nu}^{(J+3)\beta}Z_{\phi_L}^{2i}\bar{\chi}^0_{N_\alpha}
P_L\chi^0_{N_\beta}H^0_{L_i}\Big\}\nonumber\\&&\hspace{0.2cm}+H.c.
\end{eqnarray}

$\mathcal{L}^2(\nu)$ are composed of the couplings: 1. neutrino-sneutrino-MSSM neutralino;
 2. neutrino-sneutrino-lepton neutralino; 3. neutrino-slepton-chargino
\begin{eqnarray}
&&\mathcal{L}^2(\nu)=\sum_{I,J=1}^3\;\sum_{\alpha,j=1}^6\;\Big\{
-\sum_{i=1}^4\bar{\chi}_i^0\Big(
Y_\nu^{IJ}Z_N^{4i}Z_{N_\nu}^{I\alpha}Z_{\tilde{\nu}}^{(J+3)j*}P_L
\nonumber\\&&\hspace{0.4cm}+Y_\nu^{IJ*}Z_N^{4i*}Z_{N_\nu}^{(J+3)\alpha*}Z_{\tilde{\nu}}^{Ij*}P_R\Big)\chi^0_{N_\alpha} \tilde{N}^{j*}
\nonumber\\&&\hspace{0.4cm}+\sum_{i=1}^3\Big[\sqrt{2}g_LZ_{N_L}^{1i}Z_{N_\nu}^{I\alpha}
 Z_{\tilde{\nu}}^{Jj*}\delta_{IJ}-\Big((\lambda_{N^c}^{IJ}+\lambda_{N^c}^{JI})Z_{N_L}^{3i}
\nonumber\\&&\hspace{0.4cm}+\sqrt{2}g_LZ_{N_L}^{1i}\delta^{IJ}\Big)Z_{N_\nu}^{(I+3)\alpha}Z_{\tilde{\nu}}^{(J+3)j*}\Big]
\bar{\chi}^0_{Li}P_L\chi^0_{N_\alpha}
\tilde{N}^{j*}\nonumber\\&&\hspace{0.4cm}-\sum_{i=1}^2Y_\nu^{IJ}
Z_+^{2i}Z_{N_\nu}^{(J+3)\alpha}Z_{\tilde{L}}^{Ij*}\bar{\chi}^0_{N_\alpha}P_L\chi_i^+\tilde{L}_j\Big\}+H.c.
\end{eqnarray}
These neutrino couplings are in favour of the studying for neutrinos in BLMSSM.

\section{Conclusion}

In this work, we briefly introduce the major content of BLMSSM, which is the extension of MSSM with local
gauged B and L. In this model, there are new neutralinos and right handed neutrinos compared with MSSM.
We deduce the Feynman rules for neutrinos and neutralinos, and they can be used to further study neutrino masses and neutrilinos in BLMSSM.
 We also show the mass matrices of particles such as lepton neutralinos(baryon neutralinos).
 Diagonalizing the corresponding mass mixing matrices one can obtain 3 lepton neutralino (3 baryon neutralino) masses.

 At first, we deduce the new couplings of the MSSM neutralinos:
  1. MSSM neutralino-exotic lepton-exotic slepton, 2. MSSM neutralino-neutrino-sneutrino, 3. MSSM neutralino-exotic quark-exotic squark;
  Secondly, the couplings of lepton neutralinos are obtained
 from the superpotential and the interactions of gauge and matter multiplets:
 4. lepton neutralino-lepton-slepton, 5. lepton neutralino-exotic lepton-exotic slepton,
 6. lepton neutralino-lepton neutralino-lepton Higgs;
   Thirdly, the baryon neutralino couplings are deduced in the same way:
7. baryon neutralino-quark-squark, 8. baryon neutralino-exotic quark-exotic squark,
 9. baryon neutralino-baryon neutralino-baryon Higgs, 10. baryon neutralinos-$\tilde{X}$-$\psi_{X}$.
  At last, we obtain the couplings of neutrinos beyond MSSM and they are divided as：
 1. neutrino-neutrino-neutral \emph{CP}-odd Higgs; 2. neutrino-neutrino-neutral \emph{CP}-even Higgs;
3. neutrino-lepton-charged Higgs; 4. neutrino-neutrino-lepton Higgs.  5. neutrino-sneutrino-MSSM neutralino;
 6. neutrino-sneutrino-lepton neutralino; 7. neutrino-slepton-chargino.
  The obtained couplings are in favour of
 the study for neutrinos and the processes relating with neutralinos in BLMSSM.

\end{multicols}
 \end{document}